\documentclass{aastex631}

\begin{document}

\title{Discovery of Asymmetric Spike-like Structures of the 10\,au Disk around the Very Low-luminosity Protostar Embedded in the Taurus Dense Core MC~27/L1521F with ALMA}

\author[0000-0002-2062-1600]{Kazuki Tokuda}
\affiliation{Department of Earth and Planetary Sciences, Faculty of Science, Kyushu University, Nishi-ku, Fukuoka 819-0395, Japan}
\affiliation{National Astronomical Observatory of Japan, National Institutes of Natural Sciences, 2-21-1 Osawa, Mitaka, Tokyo 181-8588, Japan}

\author[0000-0002-8217-7509]{Naoto Harada}
\affiliation{Department of Earth and Planetary Sciences, Faculty of Science, Kyushu University, Nishi-ku, Fukuoka 819-0395, Japan}

\author[0000-0002-7951-1641]{Mitsuki Omura}
\affiliation{Department of Earth and Planetary Sciences, Faculty of Science, Kyushu University, Nishi-ku, Fukuoka 819-0395, Japan}

\author[0000-0002-8125-4509]{Tomoaki Matsumoto}
\affiliation{Faculty of Sustainability Studies, Hosei University, Fujimi, Chiyoda-ku, Tokyo 102-8160, Japan}

\author[0000-0001-7826-3837]{Toshikazu Onishi}
\affiliation{Department of Physics, Graduate School of Science, Osaka Metropolitan University, 1-1 Gakuen-cho, Naka-ku, Sakai, Osaka 599-8531, Japan}

\author[0000-0003-1549-6435]{Kazuya Saigo}
\affiliation{Graduate School of Science and Engineering, Kagoshima University, 1-21-40 Korimoto Kagoshima-city Kagoshima, 890-0065, Japan}

\author[0000-0001-6580-6038]{Ayumu Shoshi}
\affiliation{Department of Earth and Planetary Sciences, Faculty of Science, Kyushu University, Nishi-ku, Fukuoka 819-0395, Japan}

\author[0000-0003-4271-4901]{Shingo Nozaki}
\affiliation{Department of Earth and Planetary Sciences, Faculty of Science, Kyushu University, Nishi-ku, Fukuoka 819-0395, Japan}

\author[0000-0002-1411-5410]{Kengo Tachihara}
\affiliation{Department of Physics, Nagoya University, Furo-cho, Chikusa-ku, Nagoya 464-8601, Japan}

\author[0009-0005-4458-2908]{Naofumi Fukaya}
\affiliation{Department of Physics, Nagoya University, Furo-cho, Chikusa-ku, Nagoya 464-8601, Japan}

\author[0000-0002-8966-9856]{Yasuo Fukui}
\affiliation{Department of Physics, Nagoya University, Furo-cho, Chikusa-ku, Nagoya 464-8601, Japan}

\author[0000-0003-4366-6518]{Shu-ichiro Inutsuka}
\affiliation{Department of Physics, Nagoya University, Furo-cho, Chikusa-ku, Nagoya 464-8601, Japan}

\author[0000-0002-0963-0872]{Masahiro N. Machida}
\affiliation{Department of Earth and Planetary Sciences, Faculty of Science, Kyushu University, Nishi-ku, Fukuoka 819-0395, Japan}

\begin{abstract}

Recent Atacama Large Millimeter/submillimeter Array (ALMA) observations have revealed an increasing number of compact protostellar disks with radii of less than a few tens of au and that young Class 0/I objects have an intrinsic size diversity. To deepen our understanding of the origin of such tiny disks, we have performed the highest-resolution configuration observations with ALMA at a beam size of $\sim$0\farcs03 (4\,au) on the very low-luminosity Class~0 protostar embedded in the Taurus dense core MC~27/L1521F. The 1.3\,mm continuum measurement successfully resolved a tiny, faint ($\sim$1\,mJy) disk with a major axis length of $\sim$10\,au, one of the smallest examples in the ALMA protostellar studies. In addition, we detected spike-like components in the northeastern direction at the disk edge. Gravitational instability or other fragmentation mechanisms cannot explain the structures, given the central stellar mass of $\sim$0.2\,$M_{\odot}$ and the disk mass of $\gtrsim$10$^{-4}$\,$M_{\odot}$. Instead, we propose that these small spike structures were formed by a recent dynamic magnetic flux transport event due to interchange instability that would be favorable to occur if the parental core has a strong magnetic field. The presence of complex arc-like structures on a larger ($\sim$2000\,au) scale in the same direction as the spike structures suggests that the event was not single. Such episodic, dynamical events may play an important role in maintaining the compact nature of the protostellar disk in the complex gas envelope during the main accretion phase. 

\end{abstract}


\section{Introduction} \label{sec:intro}

Understanding the star formation process is a fundamental topic in astronomy and astrophysics. The collapse of molecular clouds leading to protostar formation \cite[e.g.,][]{Shu87,Andre00} and the subsequent interaction between stellar objects and surrounding envelopes are crucial to deciphering the mechanism to determine stellar mass \citep[e.g.,][]{Arce06,Machida_2013}. The product of the star formation process also provides initial conditions for planet formation, i.e., protoplanetary disks. Multiple physical processes, such as turbulence \cite[e.g.,][]{Padoan02,Offner08}, magnetic fields \citep[e.g.,][]{Crutcher12}, and self-gravity, are intricately coupled throughout the dynamical collapse and mass accretion phases \cite[e.g.,][]{Matsumoto11}. To observationally elucidate such processes, recent millimeter and submillimeter facilities, including interferometers offering high spatial and density dynamic range observations, have been playing effective roles.

Advances in Atacama Large Millimeter/submillimeter Array (ALMA) observations of later stages of star formation, particularly the potential planet-forming phase, have been remarkable over the past decade. The ubiquity of substructures such as rings and gaps within protoplanetary disks is increasingly evident in larger disks, typically those exceeding $\sim$50\,au in size, as highlighted by numerous high-angular-resolution observations \cite[e.g.,][]{Andrews18}. However, it is becoming clear that not all protoplanetary disks are necessarily large and massive. Survey observations across star-forming regions have shown that many disks are compact, with radii of just a few tens of au \cite[e.g.,][]{Ciza19,Long19}, and often exhibit smooth structures within the limits of their observational resolution.
The diversity of mass and size of the disk is evident even among much younger Class~0/I objects \cite[e.g.,][]{Ohashi23}. Some theoretical models tend to produce massive disks \cite[e.g.,][]{Tomida17}, and reconciling this discrepancy between observational fact and theoretical prediction is an urgent challenge (see Section~4.4 of \cite{Misugi_2023} for a recent theoretical explanation of the diversity in the disk formation). 

We focus on the dense core system MC~27 in the Taurus region ($D \sim$140\,pc; \citealt{Galli18}), where early ALMA observations with an angular resolution of $\sim$0\farcs1 revealed a compact protostellar disk with a size of $\sim$10\,au \citep{Tokuda_2017}. This system was considered to be a prestellar core before Spitzer identified a very low-luminosity object (\citealt{Bourke06,Terebey09}), whose evolutionary stage is the Class~0 phase. The dense core itself was discovered and characterized by strategically utilizing several millimeter facilities, such as the Nagoya 4\,m and the Nobeyama 45\,m telescopes \citep{Mizuno94,Onishi96,Onishi99,Onishi02}. The NH$_3$ observation by \cite{Codella97} determined the core temperature of $\sim$10\,K (L1521F in their paper). Among the dense cores in the Taurus region, it is one of the densest protostellar cores, exhibiting striking similarities to the well-known prestellar core L1544 \citep{Crapsi04} and the recently reported first core candidate, MC35-mm \citep{Fujishiro_2020,Tokuda_2020Tau} in their density profiles and deuterium fractions (see also \citealt{Tobin13}). In ALMA Cycle~0, the MC~27 system emerged as the first instance in which complex arc-like gas structures with lengths of a few thousand au were identified within a protostellar system \citep{Tokuda_2014}. Similar-sized arc structures observed at young protostellar systems have since been argued to be accretion streams \cite[e.g.,][]{Pineda_2020,Lee23}, while it remains uncertain whether the MC~27's arc and the potential streamers are physically identical or similar. Although several theories have been proposed regarding the origin of the arc structures in MC~27 (see Section~\ref{sec:MC27}), their role in the star formation process and the relationship between the peculiar structures and the compact central disk remain to be thoroughly elucidated.

From a theoretical perspective, \cite{Machida_2020} showed that starting with the magnetically subcritical state \citep{Machida18}, gravitationally collapsing cores lead to the formation of tiny protostellar disks and weak outflows due to efficient angular momentum transport by magnetic braking. In such environments with a relatively strong magnetic field, magnetic interchange instability (see also Section~\ref{sec:dis}) is likely to occur near the edge of circumstellar disks to transport the magnetic flux outward, leading to the formation of complex gas envelopes such as rings and arcs \cite[e.g.,][]{Zhao_2011}. This process provides a more dynamic direct magnetic flux removal than commonly considered steady-state magnetic field reduction, only with ohmic dissipation and ambipolar diffusion \cite[e.g.,][]{Dapp_2012,Tomida_2015}. It may offer an alternative solution (see \citealt{Tokuda23interC} and references therein) to the long-standing issue of the magnetic flux problem in star formation \citep[e.g.,][]{Nakano_1984}. Therefore, the observational characterization of tiny disk systems with complex envelopes should be an important subject to further understanding protostellar evolution and its diversity.

This paper presents an ALMA highest-resolution ($\sim$0\farcs03) 1.3\,mm continuum study of the protostellar disk in MC~27. The obtained images reveal a relatively simple structure that resolves the 10\,au disk with additional spike-like features. Incorporating a recent observational discovery, which may be relevant to a magnetic flux transport phenomenon by interchange instability \citep{Tokuda23interC}, we propose a hypothesis for forming and maintaining compact disks in a young phase of star formation. In Section~\ref{sec:obs}, we describe the observations and data reduction, then present 1.3\,mm continuum imaging results in Section~\ref{sec:res}. Section~\ref{sec:dis} integrates previously reported structures around the compact disk, discussing the star formation scenario in MC~27 and its broader implications.

\section{Observations and Data Reduction} \label{sec:obs}

We carried out ALMA Cycle~5 observations of MC~27 (P.I.: K. Tokuda, \#2017.1.00480.S) with two 12\,m array configurations, C43-9 and C43-10, using the Band~6 receivers, between 2017 October 3rd and 24. The total telescope time was $\sim$8 hours, excluding the semi-massed execution blocks. The two basebands are allocated to obtain the continuum emission at the central frequencies of 231.49 and 218.49\,GHz. The bandwidths for each setting were 2\,GHz with a channel number of 128. With the other two basebands, we targeted the molecular lines CO~($J$ = 2--1), $^{13}$CO~($J$ = 2--1), C$^{18}$O~($J$ = 2--1), and SO~($N_J$ = 5$_6$--4$_5$), but the challenging observations with the current ALMA capability could not produce high-signal-to-noise-ratio maps at a comparable resolution with the continuum data. We do not use the molecular line data hereafter, except for proving there are no hidden companion sources (Section~\ref{sec:spike}) because our immediate focus in this paper is the continuum emission of the protostar vicinity within a few au.

We used the Common Astronomy Software Application package \citep{CASAteam_2022}, v.6.5.5, in the data reduction and imaging. We employed the \texttt{tclean} algorithm with the \texttt{multi-scale} deconvolver in the analysis. 
The clean mask regions were manually defined. We experimented with several visibility weightings and found that a Briggs weighting with a robust parameter of $-$1.0 provided the best balance between resolution and sensitivity, effectively capturing the disk and its associated extra components (see Section~\ref{sec:res}). The resultant beam size and rms noise level are 0\farcs024 $\times$ 0\farcs017 with a position angle (P.A.) of 15.1\arcdeg and 20\,$\mu$Jy\,beam$^{-1}$, respectively. Note that due to the faint nature of the object, the intensity and structure somewhat depend on the imaging parameters. In Appendix~A, we present additional images with different weighting parameters (Figure~\ref{fig:contmap_change}).

\section{Results} \label{sec:res}

\begin{figure}[htbp]
    %\centering
    \flushleft
    \includegraphics[width=1.1\columnwidth]{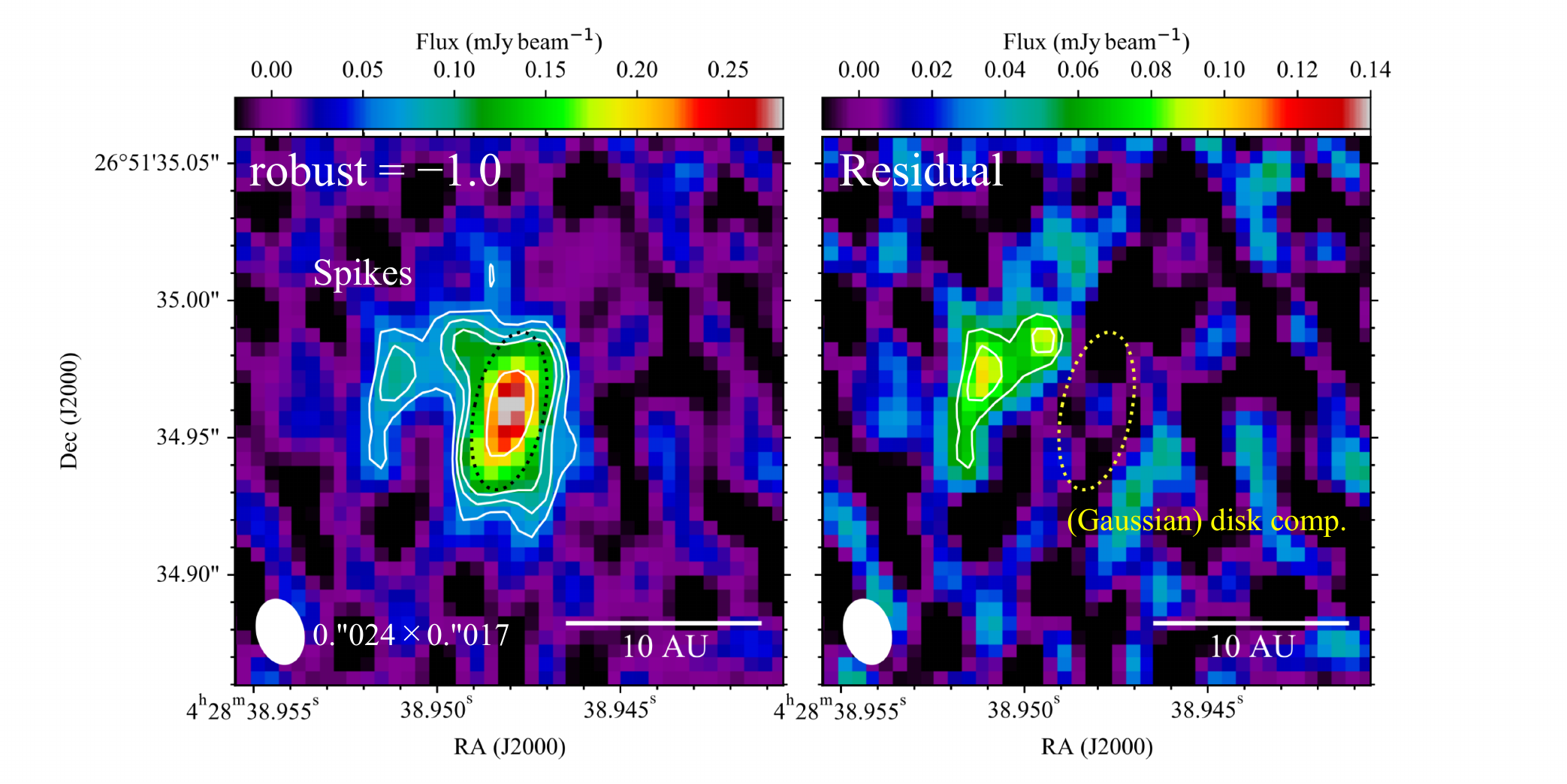}
    \caption{Left: A high-resolution 1.3\,mm continuum image toward the MC~27 protostar with a Briggs weighting at a robust parameter of $-$1 (see Section~\ref{sec:obs}). The contour levels are [3, 4, 5, 10]$\sigma$ noise levels from the lowest level. The white ellipse in the lower left corner gives the angular resolution. Right: residual image subtracting the Gaussian component in panel (a). The dotted ellipses in both panels show the fitted component whose observed major and minor FWHMs are 0\farcs058 and 0\farcs025, respectively, with a position angle of 168.5 (see the text in Section~\ref{sec:res}).}
    \label{fig:contmap}
\end{figure}

Figure~\ref{fig:contmap} (left) displays images of the 1.3\,mm continuum obtained (see Section~\ref{sec:obs}). 
The new data represent the highest-resolution ALMA images of an extremely faint protostellar disk, with a 1.3\,mm flux of $\sim$1\,mJy as the primary target. It should be noted that there are instances in cluster-forming regions where additional faint sources are reported within the field of view, separate from the main target \citep[e.g.,][]{Sharma23}.

The structure extending in a north-south direction at the center is identified as a disk component associated with the protostar. In addition to the main feature, the extra components are visible in the analyses with different imaging parameters (see Figure~\ref{fig:contmap_change}). In the case of the images with a Briggs weighting with robust parameters of $-$1 and $-$2, there are three components extending from the main source, each with emissions of $\gtrsim$3$\sigma$ significance. In the following, we refer to the central source as the $``$disk$"$ and the protruding structures that extend to the northeast as the $``$spikes.$"$

We characterize the disk and spike properties. Applying a 2D Gaussian fit to the central bright source in Figure~\ref{fig:contmap} (left) yields the observed size in FWHM of 0\farcs058 $\times$ 0\farcs025 (P.A. = 168.5$\arcdeg$), with a beam-deconvolved size of 0\farcs054 $\times$ 0\farcs018 (P.A. = 165.9$\arcdeg$).
The major axis is roughly perpendicular to the rotation axis of the disk measured in a molecular line \citep{Tokuda_2017}, suggesting that the north-south elongation is more likely to be physical, rather than a result of the beam-smearing effect. 
The disk flux is measured as $\sim$1.0\,mJy, which corresponds to 75--80\% of the total 1.3\,mm flux, 1.4\,mJy, of the detected emission over the region displayed in Figure~\ref{fig:contmap} (left). The total H$_2$ gas mass ($M_{\rm gas}$) is derived from the following equation: 
\begin{equation}
    M_{\rm gas} = \frac{F_{\rm 1.3mm} d^2}{\kappa_{\rm 1.3mm}B(T_{\rm d})},
\end{equation}
where $F_{\rm 1.3mm}$ is the total flux, $d$ is the distance to the source (140\,pc), $\kappa_{\rm 1.3mm}$ is the dust opacity per interstellar matter, $B$ is the Planck function, and $T_{\rm d}$ is the dust temperature. Assuming that the entire disk is optically thin, with $T_{\rm}$ = 20\,K and $\kappa_{\rm 1.3\,mm}$ = 0.01\,cm$^{2}$\,g$^{-1}$ \citep[e.g.,][]{Ossenkopf1994}, the $M_{\rm gas}$ is estimated to be $\sim$4 $\times$10$^{-4}$\,$M_{\odot}$.
The derived mass represents one of the lightest categories within the current ALMA survey across low-mass star-forming regions at the distance of 100--200\,pc (see \citealt{Long19,Ohashi23}). The H$_2$ column density ($N_{\rm H_2}$) is estimated using
\begin{equation}
    N_{\rm H_2} = \frac{F_{\rm 1.3mm}^{\rm beam}}{\Omega_{\rm A}\mu_{\rm H_2}m_{\rm H}\kappa_{\rm 1.3mm}B(T_{\rm d})},
\end{equation}
where $F_{\rm 1.3mm}^{\rm beam}$ is the 1.3\,mm intensity at a  position, $\Omega_{\rm A}$ is the beam solid angle, $\mu_{\rm H_2}$ is molecular weight per hydrogen molecule ($\mu_{\rm H_2}$ = 2.8), and $m_{\rm H}$ is the H atom mass. $F_{\rm 1.3mm}^{\rm beam}$ at the disk edge and peak are $\sim$0.06\,mJy\,beam$^{-1}$ and $\sim$0.28\,mJy\,beam$^{-1}$, respectively. Applying the same assumptions to derive $M_{\rm total}$, $N_{\rm H_2}$ are $\sim$5 $\times$10$^{24}$\,cm$^{-2}$ at the disk's edge and $\sim$2 $\times$10$^{25}$\,cm$^{-2}$ at the disk peak.

Next, we characterize the spike-like feature. Figure~\ref{fig:contmap} (right) shows the residual 1.3\,mm image subtracting the Gaussian component described above. This figure justifies that the observed 1.3\,mm emission having excess features in addition to the Gaussian-like component. The total flux of the excess (spikes) is $\sim$0.36\,mJy. Under the same assumptions as for the disk calculations, the measured flux corresponds to a mass of $\sim$1 $\times$10$^{-4}$\,$M_{\odot}$ and a column density of $\sim$(3--5) $\times$10$^{24}$\,cm$^{-2}$. Table~\ref{tab:disk_mass} summarizes the disk and spike properties derived in this section. 
Note that the spike structures are faint; thus, the measured properties somewhat depend on the imaging parameters and analysis methods. We have conducted the same analysis with lower-resolution data (Figure~\ref{fig:contmap_change} (left)) to outline the possible range of quantities in Table~\ref{tab:disk_mass}. Estimating additional systematic errors is challenging, but the possible uncertainties for each quantity could be a factor of $\sim$2. 
In any case, it is quite evident that there are entities with masses amounting to several tens of percent of the protostellar disk. What we present as our results may be extremely simple, but such protruding structures have been identified for the first time in the disks of young protostellar systems. 

\begin{table}[htbp]
\caption{Physical properties of the disk and spikes based on the 1.3\,mm continuum emission}
\label{tab:disk_mass}
\begin{tabular}{lccclc}
\hline
Component & Size (arcsec in FWHM)     & P.A. ($\arcdeg$)    & $F_{\rm 1.3mm}$ (mJy) & $M_{\rm gas}$ ($M_{\odot}$) & H$_2$ column density (cm$^{-2}$) \\
\hline \hline
Disk (Gaussian)     & $\sim$0.058 $\times$ 0.025 & 177.9 & 1.0--1.1               & $\sim$4 $\times$10$^{-4}$ & $\sim$(5--20) $\times$10$^{24}$  \\
Spikes (Residual)   & $\cdots$                  & $\cdots$ & 0.3--0.36               & $\sim$1 $\times$10$^{-4}$ & $\sim$(3--5)$\times$10$^{24}$        \\
\hline
\end{tabular}
\end{table}

\section{Discussion} \label{sec:dis}

\subsection{What is the origin of the spike-like structures around the disk?} \label{sec:spike}

We first discuss what the newly identified spike-like structure could be. Given its extension from the disk, it is initially presumed to be a structure that has fragmented from the disk. The enclosed mass of the protostar is estimated to be $\sim$0.2\,$M_{\odot}$ \citep{Tokuda_2017}, and the mass of the disk is extremely light at $\sim$4 $\times$10$^{-4}$\,$M_{\odot}$ (Section~\ref{sec:res}). Following a simplified formula of \cite{Kratter16}, the Toomre Q value is estimated to be $>>$1, suggesting a substantially stable state of the disk. Therefore, it is unlikely that the structure results from the disk undergoing gravitational fragmentation. Turbulent fragmentation could be a possible mechanism for forming wide binaries \citep{Offner10}, but the separation between the central disk and spikes is quite small in our case. Moreover, the available high-resolution molecular line measurements \citep{Tokuda_2017}, including the preliminary analysis of the $^{12}$CO data in the present data set, have not revealed any signs of a velocity field that would suggest the presence of additional protostellar sources, making it unlikely that the structure is a circumstellar disk of hidden companions.

Given that the known fragmentation processes mentioned above cannot explain the formation of spike-like structures, we propose magnetic interchange instability \citep{Parker_1979,Kaising_1992,Lubow_1995,Stehle_2001} as a plausible mechanism. Magnetic buoyancy acts when the magnetic field becomes strong relative to the mass loading toward the central gravitational source \citep{Lubow_1995}, and then the magnetic flux advects outward. 
Figure~\ref{fig:ii} in the Appendix illustrates the magnetic field and gas dynamics due to interchange instability based on relevant numerical studies described below. Initially, a disk and envelope system is present. 
After interchange instability occurs, magnetic flux escapes from an edge of the disk and forms a cavity within which the leaked magnetic flux and a small amount of gas expand outward, maintaining a nearly elliptical shape
\cite[e.g.,][]{Zhao_2011,Matsumoto_2017}. The surrounding gas and magnetic field from the envelope flow into the disk through channels along the cavity edge \cite[e.g.,][]{Machida_2020}.

It is possible that the spike is tracing a part of this ring structure, because the outer part farther from the protostar likely has a lower column density below the detection limit. The ring structure, due to interchange instability, grows at roughly the speed of sound (see Section~\ref{sec:MC27}), and thus it could be a recent event that occurred within the last few hundred years.

We roughly estimate the magnetic field strength conditions necessary to (1) initiate the creation of and (2) further develop interchange-instability-originated structures, such as spikes, rings, and arcs, at the edge of the protostellar disk, denoted as $B_{\rm init}$ and $B_{\rm dev}$, respectively. According to the formulae of \cite{Lubow_1995}, the condition required to trigger the instability can be described as
\begin{equation}
    B_{\rm init}^2 > \frac{2\pi\Sigma_0 G M_{*}}{r^2},
\end{equation}
where $G$ is the gravitational constant, $\Sigma_{0}$ is the surface density at the edge of the disk, and $M_{*}$ is the mass of the central star. From our observations, $\Sigma_{0}$ is approximately $\sim$5 $\times$10$^{24}$\,cm$^{-2}$ at the edge of the disk, $r$ $\sim$4\,au (Section~\ref{sec:res}). Given an $M_{*}$ of 0.2\,$M_{\odot}$ \citep{Tokuda_2017}, the required magnetic field strength $B_{\rm init}$ is on the order of $\sim$1\,G. Subsequently, for the structures formed by interchange instability to develop further, the magnetic pressure must overcome the ram pressure from the infalling envelope onto the disk. This condition is given by
\begin{equation}
    \frac{B_{\rm dev}^2}{8\pi} > \rho v^2,
\end{equation}
where $\rho$ is the density of the infalling envelope and $v$ is the velocity of the gas falling toward the disk. 
Note that we cannot use the observationally derived densities at the disk edge (Section~\ref{sec:res}) to apply as the property of infalling material, because the disk is not smoothly connected to the envelope \citep{Tsukamoto_2023PP7}. At a few au from the protostar, we adopt a value of 10$^{-13}$\,g\,cm$^{-3}$ based on numerical work \citep{Masunaga2000}. This value represents the critical density at which magnetic dissipation becomes effective, as indicated by \citep{Nakano_2002,Machida_2007}.
It is considered that the rotationally supported disks form in the region exceeding the critical density \citep{Tsukamoto_2023PP7}. A freefall velocity at $r$ = 4\,au is $\sim$9\,km\,s$^{-1}$, which is determined by the central stellar mass of 0.2\,$M_{\odot}$. In this case, the required magnetic field strength $B_{\rm dev}$ is approximately 1\,G. Therefore, the magnetic field strength necessary for the formation and subsequent development of structures due to interchange instability, though local, is strong. 
Unfortunately, determining the actual magnetic field strength through observations remains challenging, especially in a protostellar disk scale, due to some physical and technical reasons \citep[e.g,][]{Liu_2023}. 
The abovementioned field strength is slightly stronger than $\lesssim$1\,G, which is typically seen in nonideal magnetohydrodynamics simulations that involve magnetic diffusion \citep{Masson_2016,Tsukamoto_2017,Tsukamoto2023}.
However, if the parental core is penetrated by a strong average magnetic field, the gravitational contraction amplifies the magnetic field, especially on a smaller scale, creating favorable conditions for triggering interchange instability \citep{Machida_2020}. In the following, we discuss the magnetic field conditions at the envelope on the dense core scale based on the available observations.

Submillimeter continuum polarimetric observations of the MC~27 core have allowed for the determination of the magnetic field strength. Using the single-dish James Clerk Maxwell Telescope (JCMT), \cite{Soam19} and \cite{Fukaya23} detected polarized emission and estimated a plane-of-sky magnetic field strength ($B_{\rm pos}$) of 70--400\,$\mu$G based on the Davis-Chandrasekhar-Fermi method \citep{Davis51,Chand53}, providing a mass-to-flux ratio normalized by its critical value (2$\pi$$G^{1/2}$)$^{-1}$, $\mu_{\rm c}$ of $\sim$1.5--3, which is a magnetically supercritical state as a whole. We here reevaluate the magnetic field strength and mass-to-flux ratio within a 3000\,au radius from the central protostar, which is relevant to the ALMA-observed envelope-scale (see also Section~\ref{sec:MC27}). \cite{Fukaya23} reported a polarization vector dispersion of $\sim$11$\arcdeg$ from 450\,$\mu$m at the close vicinity of the protostar. The dense gas tracer N$_2$H$^{+}$ with a single dish at a similar resolution yields an FWHM velocity width of 0.28\,km\,s$^{-1}$ \citep{Tatematsu04}. The two values provide a magnetic field strength of $\sim$510\,$\mu$G using the same other assumptions \citep{Soam19,Fukaya23}. Based on the formula from \cite{Crutcher04}, $\mu_{\rm c} = 7.6 \times 10^{21} N({\rm H{_2}})/B_{\rm pos}$ is calculated to be $\sim$0.5 with a column density of 3 $\times$10$^{22}$\,cm$^{-2}$ at a radius of 3000\,au \citep{Tokuda_2016}. Note that it is not necessarily a strong constraint, but the low $\mu_{\rm c}$ value cannot exclude the central region having a strong magnetic field to realize the magnetically subcritical state. 

Note that the conditions under which interchange instability occurs are not necessarily constrained tightly, as they involve nonlinear, three-dimensional magnetohydrodynamic phenomena. Interchange instability can occur even in weaker magnetic field conditions on a dense core scale with $\mu_{\rm c}$ of $\gtrsim$2-3 \citep{Zhao_2011,Joos_2012,Machida_2014mnras}. Furthermore, the characteristics of dust grains and their growth can influence magnetic flux transport in the disk, potentially triggering interchange instability even when $\mu_{\rm c}$ is as high as  $\sim$5 \citep{Tsukamoto2023}. In any case, if interchange instability occurred in the protostellar disk of MC~27, we propose that it could explain the formation of the observed spike structures. This hypothesis can also simultaneously explain the larger structures, as discussed in Section~\ref{sec:MC27}.

In general, the onset of interchange instability depends on the resistivity evolution, which depends on the ionization degree evolution. In high-molecular-density regions such as the one studied in this article, the ionization degree is essentially determined by the dust grain-size distribution, because the surfaces of dust grains are the main recombination sites of charged particles. However, we are currently missing information on the dust grain size distribution in such a high-density region. Therefore, if our interpretation of the current observation is correct, the identification of the interchange instability may enable us to discriminate various models of resistivity in numerical simulations. Thus, further studies in this direction may provide valuable information on the evolution of the grain-size distribution, which is crucial in our understanding of planet formation.

\subsection{Summary of the MC~27/L1521F studies and a new star formation scenario}\label{sec:MC27}

The previous ALMA observations investigated MC~27 at various resolutions and molecular lines, and here we first summarize the detailed characteristics. MC~27 is a dense core with a mass of $\sim$3--4\,$M_{\odot}$, which is a general feature of similar targets in the Taurus region \citep{Onishi02}. What separates it from the others is its high central density, which is estimated to be $\sim$10$^6$\,cm$^{-3}$ at a single-dish-resolution scale, $\sim$2000--3000\,au \citep{Onishi99,Crapsi04,Kauffmann08}. At its center lies the protostar, with a very low internal luminosity of $\sim$0.07\,$L_{\odot}$, surrounded by a bipolar scattered light that spans several thousand au, likely created by an outflow activity \citep{Bourke06,Terebey09}. This bipolar nebula aligns roughly perpendicular to the north-south elongation of the central disk (Figure~\ref{fig:contmap}). 
On the other hand, the bipolar outflows observed in molecular line emission with ALMA are remarkably more compact than the scattered light nebula \citep{Tokuda_2014}, spanning only about a few hundred au, and do not connect to the central disk \citep{Tokuda_2016,Tokuda_2018}. The absence of a large outflow corresponding to the Spitzer scattered light was puzzling. Instead, arc-like structures spanning $\sim$2000\,au were observed in HCO$^+$, $^{13}$CO, and C$^{18}$O, not entirely coinciding with the edge of the scattered light \citep{Tokuda_2014,Tokuda_2018}. 

\begin{figure}[htbp]
    \centering
    %\flushleft
    \includegraphics[width=1.02\columnwidth]{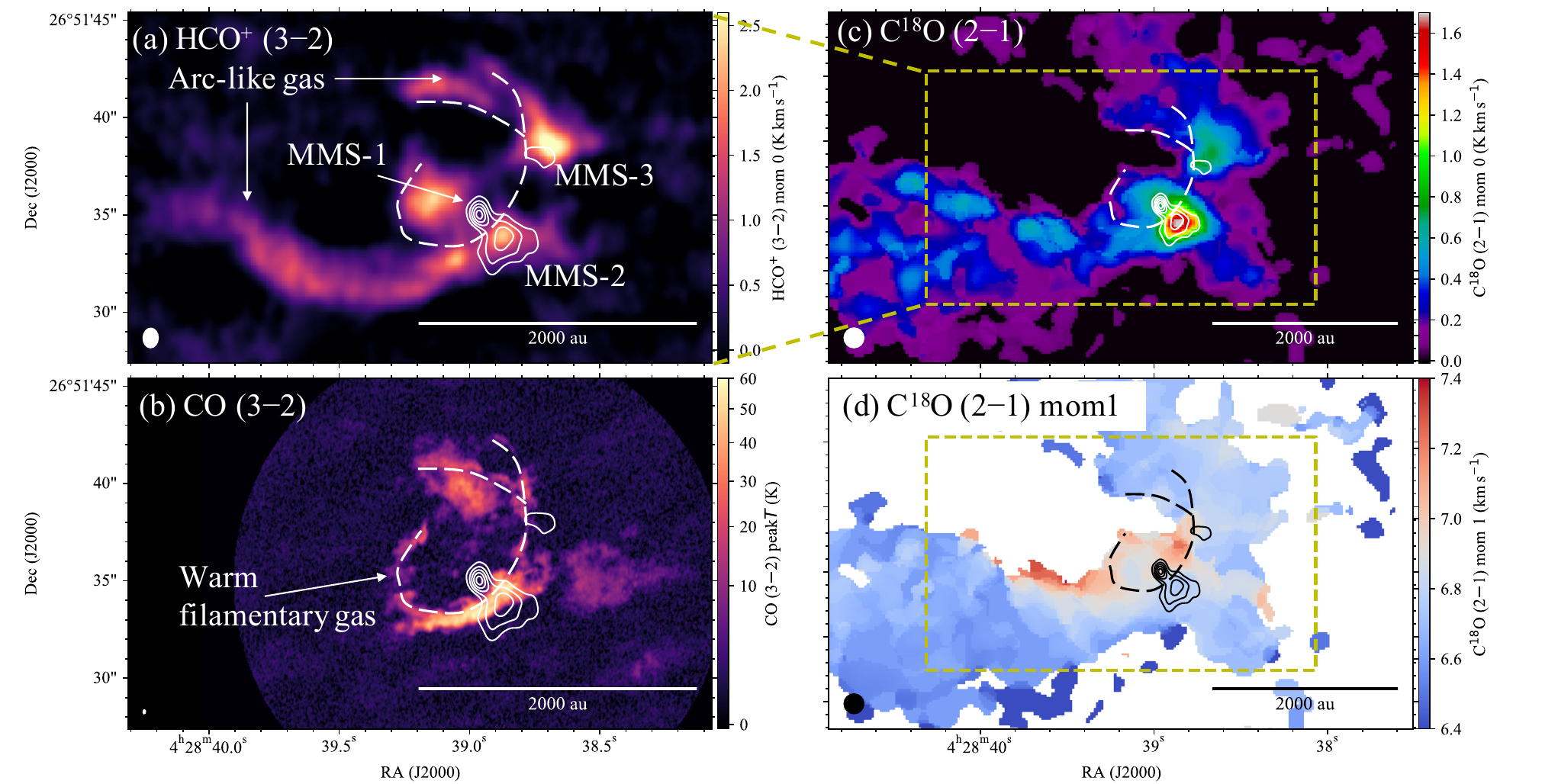}
    \caption{
    (a) Color scale image showing the HCO$^+$~(3--2) moment~0 map with a velocity range of 6.7--7.1\,km\,s$^{-1}$. The white contours show the 1.3\,mm continuum image with a contour step of 0.36\,mJy\,beam$^{-1}$ (reproduced data from \citealt{Tokuda_2014}). The contours in the other panels are the same as those in panel (a). Ellipses give the angular resolutions of each panel's image in the lower left corners.
    (b) Color scale image showing the peak brightness temperature map of CO~(3--2) with a velocity range of 0.2--6.2\,km\,s$^{-1}$ (reproduced data from \citealt{Tokuda_2018}). The white dotted lines highlight the warm filamentary gas components. The same lines are also illustrated in the other panels.
    (c) Color scale image shows the C$^{18}$O~(2--1) moment~0 map with a velocity range of 6.4--7.5\,km\,s$^{-1}$ (reproduced data from \citealt{Tokuda_2018}). 
    (d) Color scale showing the C$^{18}$O~(2--1) moment~1 map.}
    \label{fig:previousALMA}
\end{figure}

\begin{figure}[htbp]
    \centering
    %\flushleft
    \includegraphics[width=1.0\columnwidth]{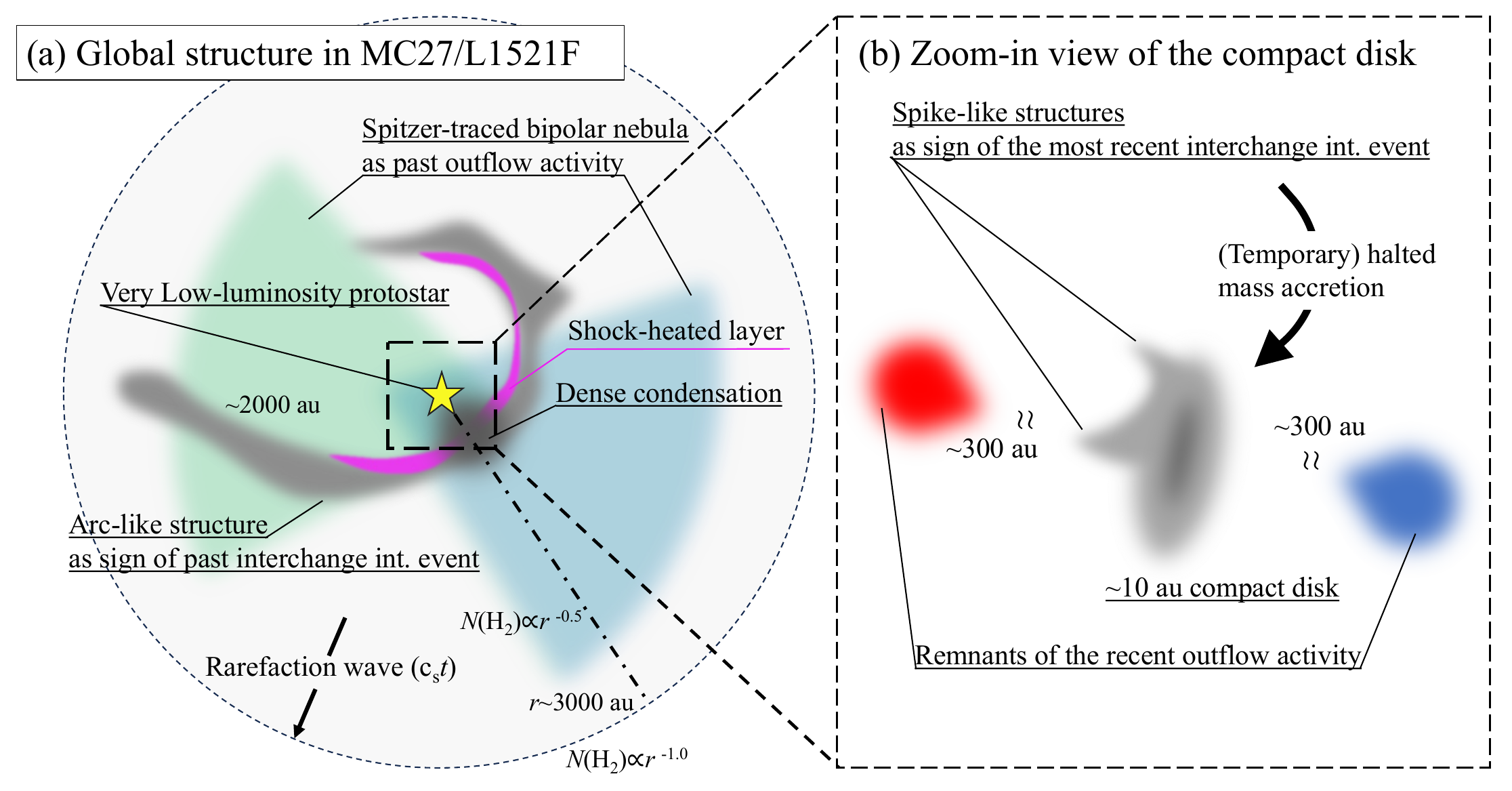}
    \caption{A schematic summary of the characteristic gas structure at the protostellar vicinity of the MC~27 dense core system.}
    \label{fig:illust}
\end{figure}

Using data from \cite{Tokuda_2014,Tokuda_2018}, Figure~\ref{fig:previousALMA} reconstructs the characteristic spatial and velocity structures of the inner $r \sim$3000 au region of the vicinity of the protostar. Panel~(a) shows the arc-like structures in HCO$^{+}$ and millimeter sources. MMS-1 is the protostellar source, which corresponds to the very low-luminosity protostar, and its disk size is just $\sim$10\,au based on our new spatially resolved image (Figure~\ref{fig:contmap}). MMS-2 and MMS-3 are starless condensations with a volume H$_2$ density of $\sim$10$^6$--10$^{7}$\,cm$^{-3}$. The particularly high-density source, MMS-2, is located in the southwest direction of the protostar. Panel~(b) shows the CO~(3--2) peak brightness temperature map. The high intensity (30--60\,K) along the filamentary cloud means that the kinematic temperature is remarkably high. The warm CO structure appears to exist inside the arc structures. Panels (c) and (d) illustrate a slightly larger scale of C$^{18}$O with a spatial filter supplement, using the Atacama Compact Array data. Beyond the HCO$^{+}$ arc structures and warm filamentary gas, there is an even more extended gas outward. Looking at the velocity structure, the southern arc seems to have a gentle velocity gradient. The systemic velocity of the dense core itself is $\sim$6.6\,km\,s$^{-1}$ as measured by the single-dish studies \citep{Onishi99,Tatematsu04}, and thus the bulk velocity of the arcs and their surroundings is the same as the parental core. The inner side, closer to the protostar, appears to be redshifted. The above features are also schematically summarized in Figure~\ref{fig:illust}, which is discussed in more detail later.

For the star formation scenario in MC~27, particularly the origin of the distinctive arc structures, we initially proposed gravitational torques resulting from interactions among multiple protostellar sources, which were believed to be hidden in the dense condensations MMS-2 and MMS-3 \citep{Tokuda_2014,Matsumoto_2015}. However, subsequent observations, including the present highest-resolution study, have clarified that no such point sources likely exist within the dense sources. In addition, as we discussed in Section~\ref{sec:spike}, the dense region of the MC~27 core may have a strong magnetic field ($\mu_c$ $\sim$0.5), which could effectively suppress the fragmentation and subsequent formation of binary and multiple stellar systems \citep{Machida_2008binary,Hennebelle2008,Wurster2019}. This finding may exclude the multiple system-driven scenarios as an origins of the arc-like structure, although another promising system has been discovered \citep{Lee23}. \cite{Tokuda_2017,Tokuda_2018} argued that the small ($\sim$10\,au) size of the protostellar disk was also once attributed to dynamical truncation by multiple components, but this possibility can be similarly dismissed. The presence of possible shock-heated gas, as observed in CO~(3--2), located around the disk, led us to propose turbulent stripping as truncating the disk. A colliding gas flow onto a face-on disk can truncate the outer edge of the protostellar disk, but it requires a high-velocity ($\sim$20\,km\,s$^{-1}$) and dense (5 $\times$10$^{6}$\,cm$^{-3}$) stream \citep{Wijnen16}. Such extreme conditions are unlikely in the low-mass dense core, necessitating the reconsideration of the overall description of the star formation scenario in this particular system.

We discuss the formation of the large-scale arc structures and the newly detected spikes in MC~27 through magnetic interchange instability (see Section~\ref{sec:spike}). Unlike the previously mentioned scenarios, magnetic interchange instability does not impose physically inconvenient conditions, such as a highly turbulent environment and/or multiple protostellar interactions. Magnetic interchange instability allows for the creation of asymmetric structures around the disk \citep{Zhao_2011, Matsumoto_2017} and efficient angular momentum transportation through magnetic braking. Because the disk is the primary outflow driver due to magnetocentrifugal force \cite[e.g,][]{Machida_2013,Commerson22}, interchange instability disrupts the accretion disk and may terminate the outflow activity as well as the accretion, at least temporarily.

Figure~\ref{fig:illust} summarizes the large- and small-scale picture of MC~27. Our newly proposed star formation scenario for MC~27 is as follows. Initially, just after the protostar formation, the first episode of interchange instability occurrs. A parcel of gas at the edge of the disk begins to leak out from the disk into the infalling or circumstellar envelope, accompanying a large amount of the magnetic flux, leading to the formation of ring-like structures or arcs as parts of incomplete ring features. 
If the rings and arcs were formed due to interchange instability, the interior would be a cavity with low gas density but a strong magnetic field. %, where gas dynamical time is determined by Alfv\'en speed, exceeding the sound speed. 
The leaked gas collides with the circumstellar material, and shocks should occur at the (inner) edge of the rings and arcs, which is consistent with the observed characteristics of higher temperatures inside the arcs (Figures~\ref{fig:previousALMA} and \ref{fig:illust}) and the partial detection of shock-related tracers such as SO and methanol \citep{Favre_2020}. Meanwhile, the arcs themselves expand outward on the order of the sound speed, $c_{\rm s}$ \citep{Zhao_2011,Matsumoto_2017}. If slight density asymmetries existed during the protostar formation, the interchange-instability-driven arc would expand toward a specific direction \citep{Krasnopolsky_2012,Tokuda23interC}. Assuming their expansion speed as $c_{s}$, 0.2\,km\,s$^{-1}$ at 10\,K, the dynamical time for these structures would be $\sim$5 $\times$10$^5$\,years (=2000\,au/0.2\,km\,s$^{-1}$). If we assume that the small-scale spike-like structures with a few au, developed at a similar speed, we are likely observing a very recent event, on the order of 100 years. If we interpret the former time, $\sim$5 $\times$10$^5$\,years, as the same as the mass accretion history, growing to the currently identified stellar mass of 0.2\,$M_{\odot}$, the time-averaged accretion rate would be $\sim$4 $\times$10$^{-6}$\,$M_{\odot}$\,yr$^{-1}$. This is consistent with theoretical predictions just after the protostar formation \cite[e.g.,][]{Hunter77,Shu77,Whitworth85,Tomisaka96}. The rarefaction wave could also spread outward on a similar timescale, $c_{\rm s}t$ \citep{Shu77}, potentially explaining the currently observed two-power-law H$_2$ column density, $N$(H$_2$), profile in the MC~27 system, which shows a shallower slope than the outer ($\gtrsim$3000\,au) part \citep{Tokuda_2016}. 
The rarefaction wave is thought to spread just after forming the first hydrostatic core \citep{Larson69}. Because interchange instability occurs at the protostellar disk edge due to magnetic diffusion \citep{Machida_2020}, the relevant structure growth must begin at least after the disk formation, i.e., the second collapse. The presence of the arcs inside the radius where the column density profile transits, i.e., the front of the rarefaction wave (Figure~\ref{fig:illust}a), does not contradict the theoretical expectations described above.

During the active mass accretion phase, a powerful outflow likely formed, creating the cavity traced by Spitzer's scattered light. If the mass accretion and ejection were continuous, corresponding components should be observable in the molecular line emission, but their absence suggests that the outflow has been at least temporarily halted. Indeed, the current low luminosity indicates an almost negligible mass accretion rate of $\lesssim$ 2 $\times$10$^{-8}$\,$M_{\odot}$ \citep{Tokuda_2017}. The most recent interchange instability event could have disrupted the disk's asymmetry and temporarily quenched the outflow (e.g., Figure~7c in \citealt{Machida_2020}), and we see the compact outflowing components as the remnants in this case (Figure~\ref{fig:illust}(b)). The directions of the large-scale arc and the disk's spike are consistent with each other, which indicates the interchange-instability-driven structure cannot expand toward the higher-density region around MMS-2. In summary, we suggest that at least two times episodes of drastic magnetic field leakage due to interchange instability realize the complex gas distribution and maintain the disk size as quite small, $\sim$10\,au. It should be noted that \cite{Matsumoto_2017} and \cite{Machida_2020} have already proposed the origin of MC~27's large-scale arc structures as being due to interchange instability. By combining the current observational evidence, including the ALMA highest-resolution image, we reconstruct and consistently explain the overall star-formation scenario for the first time.

\subsection{Implications for the broader picture of star and protostellar disk formation}\label{sec:Inp}

While MC~27 presents a highly peculiar system, it is crucial to examine whether the star formation scenario that we devise here in Section~\ref{sec:MC27} is generally applicable. Prior to this MC~27 work, a ring structure with a diameter of $\sim$7000\,au possibly driven by interchange instability was reported in the Corona Australis (CrA) IRS~2 Class~I system \citep{Tokuda23interC}. Unlike MC~27, the single ring feature suggests a solitary interchange instability event. The CrA IRS~2 system, formed at the edge of a dense cluster-forming complex, likely did not experience external disturbances during the ring's growth phase, preserving a relatively pristine shape. However, conditions conducive to interchange instability could be frequently realized at the disk's edge \citep{Machida_2020}, and, indeed, theoretical calculations often reveal multiple ring structures \citep[e.g,][]{Stehle_2001}, which are also likely to be disturbed by gas accretion and outflow.

Interchange instability also allows for the formation of asymmetric arc structures in the protostellar envelope without initially imposing strong turbulence (see \citealt{Offner08}). 
These arc-like structures, as incomplete rings, probably correspond to streamers in numerous protostellar targets. While further individual studies are necessary to discuss the origins of such structures observationally, arc-like structures or rings spanning a few thousand astronomical units, similar to those discovered in MC~27, have been frequently identified in Class~0/I targets \citep[e.g.,][]{Alves_2020,Murillo_2022,Mercimek_2023,Sai_2023,Sharma23}. The possibility that these are parts of rings formed through interchange instability cannot be dismissed in favor of a simple accretion interpretation. If this is the case, the disks formed in a strongly magnetized condition that drives the interchange instability, as in MC~27, are expected to remain small without showing a powerful outflow. \cite{Aso19} presented some Class~0 objects with extensive envelopes but no detectable outflows, suggesting that the protostars may have formed from magnetically subcritical cores (see the discussion in \citealt{Machida_2020}). High-resolution observations targeting small disks to search for spike structures and exploring the relationship between disk size and the presence or absence of a surrounding arc-like envelope, as well as outflow activities, will be beneficial to gain observational insights into indirect pieces of evidence for the role of the magnetic field in star formation and subsequent protoplanetary disk formation. This type of study can potentially offer insights into the magnetic flux problem in the star-formation process.

\section{Summary}\label{sec:summary}

We have conducted the highest-resolution ($\sim$0\farcs03) ALMA observations at a wavelength of 1.3\,mm of the puzzling dense core MC~27/L1521F in Taurus, which harbors a very low-luminosity protostar. This measurement marks the first instance of executing such a high-resolution study on a faint protostellar source with a 1.3\,mm flux of $\sim$1\,mJy. Our results spatially resolved the 10\,au disk and captured spike-like features extending northeastward from the disk. We propose that this faint feature could be the result of a recent event involving magnetic flux advection, because other possible mechanisms, such as gravitational instability and turbulent fragmentation, cannot explain the new feature. The dynamic magnetic flux transport events (interchange instability) occurring in this system are not single; at least one previous event has played a crucial role in shaping the compact disk of this system and making the large-scale arc-like feature, which resembles recently reported accretion streamers. We suggest that the interchange instability might be at work in other protostellar systems hosting compact ($\lesssim$10\,au) disks surrounded by arc- or ring-like structures. By investigating the history of magnetic flux transport imprinted in complex envelopes, we can deepen our understanding of the diversity of star formation, particularly the origin of small-sized disks, and subsequent planet formation as inferred from grain-size distributions that satisfy the conditions for the occurrence of interchange instability.

%\begin{itemize}
%\item[1.] hoge
%\end{itemize}

\begin{acknowledgments}

We would like to thank the anonymous referee for useful comments that improved the manuscript. This paper makes use of the following ALMA data: ADS/JAO.ALMA\#2011.0.00611.S, 2015.1.00340.S, and 2017.1.00480.S. ALMA is a partnership of ESO (representing its member states), the NSF (USA), and NINS (Japan), together with the NRC (Canada), MOST, and ASIAA (Taiwan), and KASI (Republic of Korea), in cooperation with the Republic of Chile. The Joint ALMA Observatory is operated by the ESO, AUI/NRAO, and NAOJ. This work was supported by an NAOJ ALMA Scientific Research grant (Nos. 2022-22B) and grants-in-aid for scientific research (KAKENHI) of the Japan Society for the Promotion of Science (JSPS; grant Nos. JP18H05436, JP18H05437, JP20H01945, JP20H05645, JP21H00046, JP21H00049, JP21K03617, JP21K13962, JP23K03464, and JP23H00129). The main content observations for this study were conducted more than 6 years before the writing of this paper. We have long regretted not being able to release these valuable ALMA data (2017.1.00480.S) in a publication format, but the serendipitous discovery of a large-scale gas ring in the Corona Australis region \citep{Tokuda23interC} provided a breakthrough for our astronomical interpretations in this study. We would like to express our gratitude once again to the objects, including MC~27, that have shown us the dynamic aspects of star formation.
\end{acknowledgments}

\appendix

\section{1.3\,mm continuum imaging with different CLEAN weighting parameters}

\renewcommand{\thefigure}{A\arabic{figure}}
\setcounter{figure}{0} 

\begin{figure}[htbp]
    %\centering
    \flushleft
    \includegraphics[width=1.1\columnwidth]{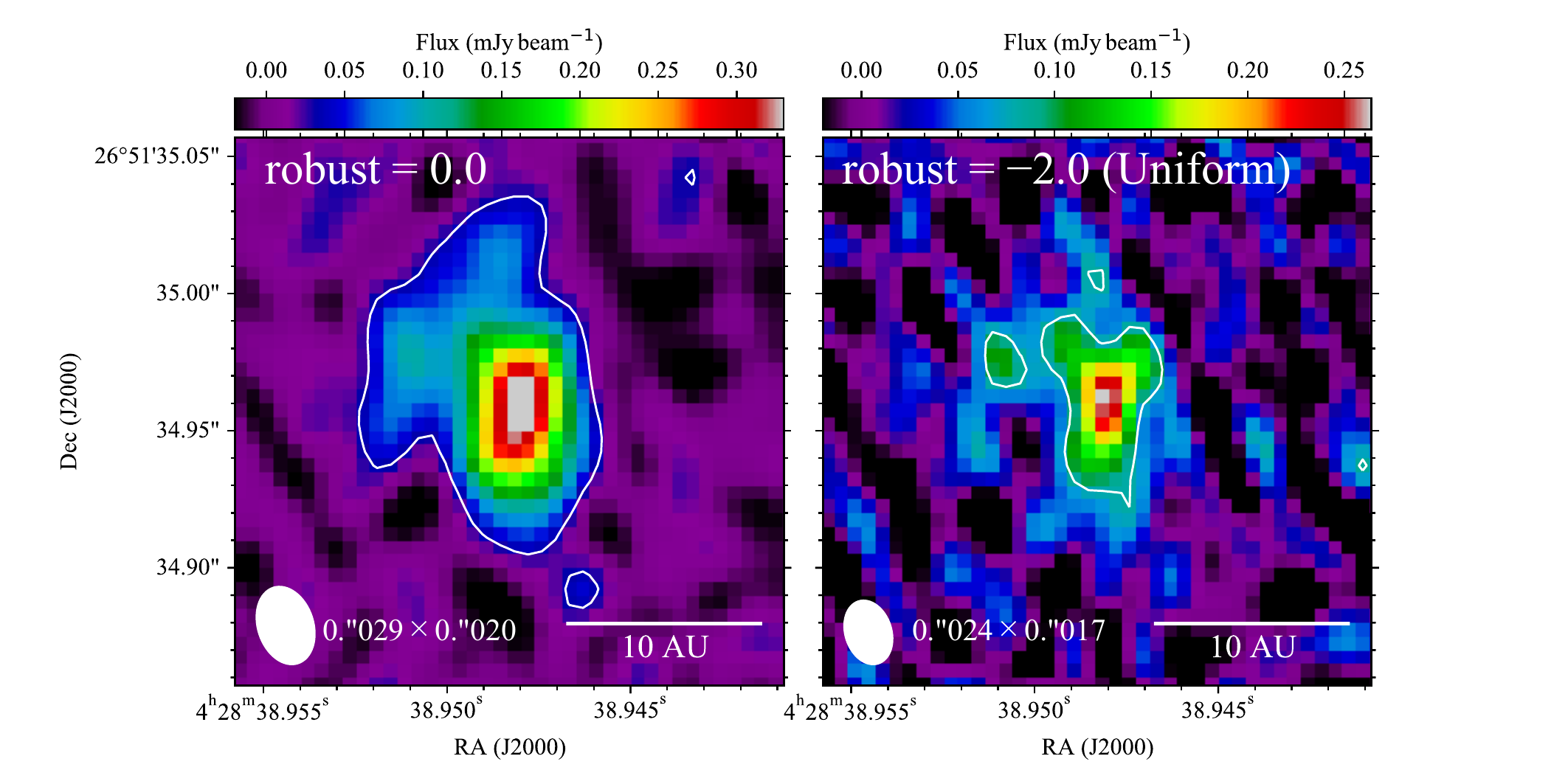}
    \caption{Left: a high-resolution 1.3\,mm continuum image of toward the MC~27 protostar with the Briggs weighting at a robust parameter of 0.5 (see Section~\ref{sec:obs}). The contour shows three times the noise level of the continuum image. The angular resolution is given by the white ellipse in the lower left corner. Right: the same as the left panel, but for the Uniform weighting for the imaging.}
    \label{fig:contmap_change}
\end{figure}

Figure~\ref{fig:contmap_change} presents two imaging results with different Briggs weighting parameters, robust = 0.0 and $-$2.0 (Uniform), to ensure the fidelity of the faint features of the disk and excess, i.e., spike structures. The resultant beam sizes of the Briggs and Uniform images are 0\farcs029 $\times$ 0\farcs020 (P.A. = 18.0\arcdeg) and 0\farcs024 $\times$ 0\farcs017 (P.A. = 17.6\arcdeg), respectively. The rms noise levels are $\sim$11\,$\mu$Jy\,beam$^{-1}$ (robust = 0.0) and $\sim$27\,$\mu$Jy\,beam$^{-1}$ (Uniform). 
Typically, Uniform weighting is expected to enhance angular resolution, but in this instance, it resulted in a resolution comparable to that obtained with a robust parameter of $-$1.0 (as illustrated in Figure~\ref{fig:contmap}). The use of a robust parameter of $-$1.0 provided the advantage of higher sensitivity in our data.

\section{A Schematic view of interchange instability}
\renewcommand{\thefigure}{B\arabic{figure}}
\setcounter{figure}{0} 

\begin{figure}[htbp]
    \centering
    %\flushleft
    \includegraphics[width=0.8\columnwidth]{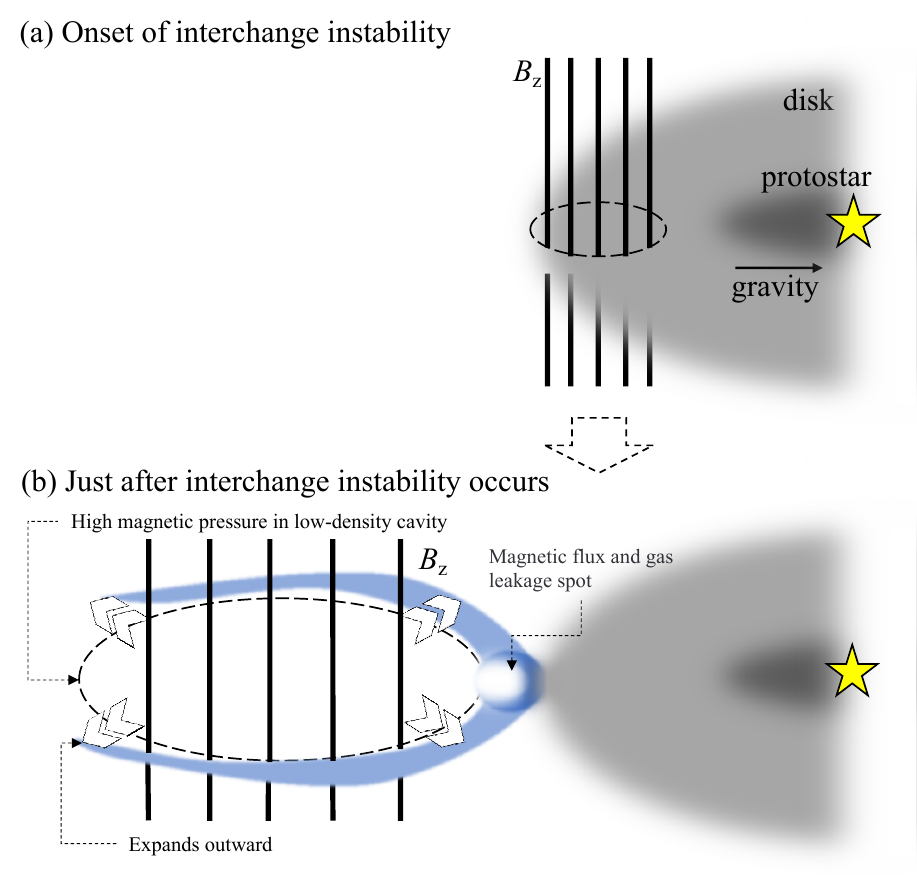}
    \caption{A schematic view of the magnetic field and gas distributions due to interchange instability around a protostellar disk.}
    \label{fig:ii}
\end{figure}

Figure~\ref{fig:ii} illustrates the schematic view of interchange instability around the disk.

\software{astropy \citep{Astropy18}, CASA \citep{CASAteam_2022}}
%Cloudy \citep{2013RMxAA..49..137F}, 
%Source Extractor \citep{1996A&AS..117..393B}}

%\bibliographystyle{aasjournal}



\end{document}